\documentclass[
 reprint,
superscriptaddress,
 amsmath,amssymb,
 aps,
 pra,
floatfix,
]{revtex4-2}

\usepackage{graphicx}
\usepackage{dcolumn}
\usepackage{bm}

\usepackage{soul}
\usepackage{braket}
\usepackage[colorlinks=true,urlcolor=blue,citecolor=blue,linkcolor=blue]{hyperref}
\usepackage[dvipsnames]{xcolor}
\usepackage{mathtools}
\usepackage{array}
\usepackage{enumerate}   
\usepackage{siunitx}

\DeclareMathOperator{\pump}{\mathrm{p}}
\DeclareMathOperator{\signal}{\mathrm{s}}
\DeclareMathOperator{\idler}{\mathrm{i}}

\begin{document}

\preprint{APS/123-QED}

\title{High-dimensional maximally entangled photon pairs in parametric down-conversion}

\author{Richard Bernecker}
 \email{richard.bernecker@uni-jena.de}
\affiliation{%
 Theoretisch-Physikalisches Institut, Friedrich-Schiller-Universität Jena, Max-Wien-Platz 1, D-07743 Jena, Germany}
\affiliation{
 Helmholtz-Institut Jena, Fröbelstieg 3, D-07743 Jena, Germany 
}
 
\author{Baghdasar Baghdasaryan}
\affiliation{%
Institut für Angewandte Physik, Friedrich-Schiller-Universität Jena, Albert-Einstein-Str. 6, D-07745 Jena, Germany
}

\author{Stephan Fritzsche}
\affiliation{%
 Theoretisch-Physikalisches Institut, Friedrich-Schiller-Universität Jena, Max-Wien-Platz 1, D-07743 Jena, Germany}
\affiliation{
 Helmholtz-Institut Jena, Fröbelstieg 3, D-07743 Jena, Germany 
}

\date{\today}

\begin{abstract}
Photon pairs generated from spontaneous parametric down-conversion are a well-established method to realize entangled bipartite photonic systems. Laguerre-Gaussian modes, which carry orbital angular momentum (OAM), are commonly exploited to engineer high-dimensional entangled quantum states. 
For Hilbert spaces with dimension $d>2$, maximally entangled states (MESs) help to improve the capacity and security of quantum communication protocols, among several other promising features. However, the direct generation of MES in well-defined high-dimensional subspaces of the infinite OAM basis has remained a challenge. Here, we formalize how the spatial distribution of the pump beam and the nonlinear profile of the crystal can be simultaneously
utilized to generate MES without additional spatial filtering of OAM modes within a subspace. We illustrate our approach with maximally entangled qutrits ($d=3$) and ququints ($d=5$). 
\end{abstract}

\maketitle

\newpage
\section{\label{sec:Introduction}Introduction}
The generation and control of entangled photons have evolved into an indispensable tool for the extensive research and development of quantum technologies in branches like quantum cryptography, quantum computation, and quantum imaging \cite{moreau2019imaging, magana2019quantum, sidhu2021advances, zhong2020quantum}. Great interest has focused on the entanglement between high-dimensional quantum systems with dimensionality $d>2$, so-called \textit{qudits} \cite{erhard2020advances}. In comparison to the frequently explored qubits ($d=2$), qudits provide higher information capacities and improved security for quantum communication protocols \cite{bechmann2000quantum, cerf2002security, walborn2006quantum} or a notable increase of efficiency for quantum-computational tasks \cite{lanyon2009simplifying, wang2020qudits}. 

An important class of bipartite states are maximally entangled states (MESs) like the famous Bell states for $d=2$. MESs are an ideal target for applications in high-dimensional systems because of the complete non-separability of the two distinct subsystems and perfect quantum correlations within the total system \cite{nielsen2010quantum}.

High-dimensional entangled quantum states can be realized with regard to different degrees of freedom (DOFs) of light, as shown in polarization \cite{bogdanov2004qutrit, fedorov2011entanglement, sekga2023high}, frequency \cite{ramelow2009discrete, bernhard2013shaping, kues2017chip}, temporal modes \cite{donohue2013coherent, grassani2015micrometer, martin2017quantifying}, path identity \cite{kysela2020path}, and photon number \cite{bimbard2010quantum}. Much attention has been devoted to orbital angular momentum (OAM) states of light \cite{groblacher2006experimental, krenn2017orbital, sit2017high} due to facile scalability in dimensions \cite{krenn2014generation, fickler2016quantum} and their application for long-distance quantum communication in free space or through mode fibers \cite{krenn2015twisted, willner2015optical}.

Spontaneous parametric down-conversion (SPDC) is widely employed to 
generate entangled bipartite systems \cite{anwar2021entangled}. In SPDC, a photon pair or \textit{biphoton} is emitted after the interaction of a pump photon with a crystal of second-order nonlinearity. The generated photon pairs intrinsically possess high-dimensional entanglement in OAM, which has been demonstrated experimentally \cite{mair2001entanglement, fickler2012quantum}. In particular, the spatial entanglement between the photon pairs is distributed over many OAM modes, also known as spiral bandwidth \cite{torres2003quantum, di2010measurement}. The amplitudes of the OAM modes within the spiral bandwidth must be uniformly distributed to ensure maximal entanglement.

However, the step from high-dimensional entangled states to MESs is difficult to realize in experiments \cite{wang2017generation}. Until the present, post-selection procedures like "Procrustean filtering" have been applied to equalize the OAM amplitudes of the biphoton state \cite{vaziri2003concentration, dada2011experimental, chen2020coherent}, but limit the number of usable photon pairs and often require additional steps in the set-up. An open question concerns the efficient generation of a MES from SPDC within a well-defined subspace of the infinite OAM space.

Spatially engineered pump beams have been a popular way to shape the OAM amplitudes of the biphoton state \cite{liu2020increasing, baghdasaryan2020enhanced}. For instance, an optimal weighted superposition of Laguerre-Gaussian modes \cite{torres2003preparation, kovlakov2018quantum, liu2018coherent, bornman2021optimal}  can equalize the amplitude of desired OAM modes. However, this approach does not automatically lead to the generation of MESs, due to the presence of additionally created OAM modes through the engineered pump beam.

Another well-known technique to manipulate the biphoton state concerns the use of tailored nonlinear crystals. Periodically poled crystals \cite{torres2004quasi} with a sinc-like phase-matching function (PMF) are favored to optimize the photon-pair source brightness. Recently, innovative domain-engineering techniques \cite{branczyk2011engineered, dixon2013spectral, tambasco2016domain, graffitti2017pure} have introduced alternative PMFs like in Gaussian shape \cite{pickston2021optimised, dosseva2016shaping}, which minimize the spectral and spatial correlations between the photon pairs without post-filtering \cite{graffitti2018independent, baghdasaryan2023enhancing}. So far, \textit{customized} PMFs have not been investigated for generating maximally entangled photon pairs. 

In this theoretical work, we aim to combine the pump beam and crystal engineering techniques in order to maximize the entanglement between photon pairs. We formulate conditions for the spatial profile of the pump and the PMF of the crystal that optimize the generation of desired high-dimensional MESs. Our method eliminates the need for spatial post-selection in the OAM subspace. 

\section{\label{sec:Theoretical Foundations}Theoretical Foundations}

Spontaneous-parametric down-conversion is a nonlinear optical process in which correlated photon pairs are created under energy conservation after a high-intensity pump beam illuminates a crystal with second-order susceptibility $\chi^{(2)}$. The simultaneously emitted photons of a pair are called \textit{signal} and \textit{idler}.  The momentum conservation $\bm{k}_{\pump}=\bm{k}_{\signal}+\bm{k}_{\idler}$ ensures the constructive interference between the propagating waves inside the crystal. This wave-vector equation with $\bm{k}_j$ defines the so-called phase-matching condition, where $j \in \{ \pump, \signal, \idler\}$ refer to pump, signal, and idler, respectively.

\subsection{\label{subsec:Biphoton state mode function}Geometry and assumptions}

In our study, we follow typical quantum optical assumptions for the SPDC process to obtain the bipartite quantum state of signal and idler, the so-called biphoton state. 

Since the intensity of the pump beam exceeds the photon pair intensity several orders of magnitude, we assume the pump as undepleted and neglect further quantum effects. We fix the phase-matching relation as type II-SPDC \cite{kurtsiefer2001high}, such that signal and idler have orthogonal polarization. We choose the pump beam to propagate along the $z$-axis focused in the center of the crystal at $z=0$. The created signal and idler photons are assumed to propagate almost parallel (quasi-collinear) to the pump direction. Since most optical elements only support paraxial waves, we apply the paraxial approximation $|\bm{q}| \ll |\bm{k}|$ to separate the wave vector $\bm{k}$ into a transverse momentum vector $\bm{q}$ and the longitudinal component 
\begin{equation}
    k_z = \sqrt{k(\omega)^2-|\bm{q}|^2} \approx k(\omega) - \frac{|\bm{q}|^2}{2 k(\omega)},
    \label{ParaxialApprox}
\end{equation}
where the wave number is given by $k=\frac{n \omega}{c}$. In this regime $\bm{q}$ is referred to as the \textit{spatial} and $\omega$ as the \textit{spectral} DOF.
Furthermore, when the transverse size of the crystal is large, compared to the pump beam waist, the conservation of the transverse momentum $\bm{q}_{\pump}=\bm{q}_{\signal} + \bm{q}_{\idler}$ is fulfilled.
In our framework, all photons are assumed to be monochromatic and to fulfill the energy conservation $\omega_{\pump} = \omega_{\signal} + \omega_{\idler}$ exactly. This condition is experimentally enforced with frequency filters in the set-up and is known as the narrowband regime. These approximations account for a broad class of experimental scenarios that are based on the entanglement of OAM light modes \cite{groblacher2006experimental, walborn2010spatial, palacios2011flux, krenn2014generation, sevilla2024spectral}. 

Under the mentioned assumptions and approximations, the biphoton state of a signal and idler pair with well-defined polarization and frequencies in momentum representation reads \cite{walborn2010spatial, baghdasaryan2022generalized}
\begin{eqnarray}
     \ket{\psi} = \mathcal{N}  && \iint d\bm{q}_{\signal} \: d\bm{q}_{\idler} \: V_{\pump}(\bm{q}_{\signal}+\bm{q}_{\idler}) \: \Phi_{\text{PM}}(\Delta k_z) \nonumber \\ 
     && \times \: \hat{a}_{\signal}^{\dagger}(\bm{q}_{\signal}) \: \otimes \: \hat{a} _{\idler}^{\dagger}(\bm{q}_{\idler}) \ket{\text{vac}},
     \label{GeneralBiphotonstate}   
\end{eqnarray}
where $\mathcal{N}$ is a normalization constant and $\Delta k_z = k_{\pump,z} - k_{\signal,z} - k_{\idler,z} $ the longitudinal wave vector mismatch. Evidently, the state is obtained by the integration over all possible spatial states of signal and idler generated from the vacuum by creation operators $\hat{a}^{\dagger}(\bm{q})$. 

The angular spectrum $V_{\pump}(\bm{q}_{\pump})$ gives rise to the transverse spatial distribution of the pump beam, whereas the phase-matching function (PMF) $\Phi_{\text{PM}}(\Delta k_z)$ depends on the crystal structure. The phase mismatch along the $z$-axis follows from Eq. \eqref{ParaxialApprox} as 
\begin{small}
\begin{eqnarray}
    \Delta k_z &&= k_{\pump}(\omega_{\signal}+\omega_{\idler}) - k_{\signal}(\omega_{\signal}) - k_{\idler}(\omega_{\idler}) \nonumber \\
    &&- \frac{|\bm{q}_{\signal}+\bm{q}_{\idler}|^2}{2k_{\pump}(\omega_{\signal}+\omega_{\idler})}  - \frac{|\bm{q}_{\signal}|^2}{2k_{\signal}(\omega_{\signal})}  - \frac{|\bm{q}_{\idler}|^2}{2k_{\idler}(\omega_{\idler})}.
\end{eqnarray}
\end{small}
The minimization of $\Delta k_z $ guarantees an efficient down-conversion process and is realized by techniques like angle tuning in birefringent crystals \cite{karan2020phase} or by quasi-phase-matched nonlinear crystals \cite{torres2004quasi}. The expression of the PMF in Eq. \eqref{GeneralBiphotonstate} is similar to a Fourier transform of the nonlinear susceptibility profile $\chi^{(2)}(z)$ \cite{dosseva2016shaping, graffitti2017pure}, which we assume to be a general function along the $z$-axis (i.e. an one-dimensional profile):
\begin{equation}
    \Phi_{\text{PM}}(\Delta k_z) = \int_{-\frac{L}{2}}^{\frac{L}{2}} dz \:  \chi^{(2)}(z) \: \mathrm{e}^{ i \Delta k_z z},
    \label{Phase-matching-Function}
\end{equation}
where $L$ is the length of the crystal. 

\subsection{\label{subsec:Decomposition in LG basis}Decomposition in LG basis}

The discussion on maximizing the entanglement in OAM by shaping the pump requires a general spatial profile for the pump beam. Within the paraxial regime, any transverse profile can be decomposed into the Laguerre Gaussian (LG) basis since they form a complete and orthogonal set of optical modes. Therefore, we initially model the pump beam as an LG mode
\begin{equation}
   V_{\pump}(\bm{q}_{\pump}) = \text{LG}^{\ell_{\pump}}_{p_{\pump}}(\bm{q}_{\pump},w_{\pump}),
    \label{PumpinLG}
\end{equation}
and later generalize it to a superposition of LG beams. The choice of LG modes is particularly advantageous because they are eigenstates of the orbital angular momentum operator, enabling easy control over the OAM amplitudes of the biphoton state. LG beams used for SPDC are mostly generated with hologram-displaying spatial light modulators (SLMs) \cite{bolduc2013exact}.

The angular spectrum of a LG mode in the momentum space at $z=0$ is given by
\begin{eqnarray}
    \text{LG}^{\ell}_{p}(\bm{q},w) && = \sqrt{\frac{w^2 p!}{2\pi(p + |\ell|)!}} (-1)^{p+\frac{\ell}{2}} \left( \frac{|\bm{q}|}{\sqrt{2} w} \right)^{|\ell|} \nonumber \\ && \times \mathrm{e}^{-\frac{|\bm{q}|^2 w^2}{4}} \: \text{L}_{p}^{|\ell |}\left(\frac{|\bm{q}|^2 w^2}{2}\right) \mathrm{e}^{i \ell \text{Arg}(\bm{q}) },
\end{eqnarray}
where $\text{L}_{p}^{|\ell|} \left( \cdot \right)$ are the generalized Laguerre polynomials and the continuous parameter $w$ is the beam waist.
The radial index $p \in \mathbb{N}_0$ refers to the radial nodes observed in the transverse plane. The spiral index $\ell  \in \mathbb{Z}$ characterizes the amount of OAM (projected on the z-axis, in units of $\hbar$) each photon carries. $| \ell |$ also dictates how often the phase changes from $-\pi$ to $\pi$ on a closed path around the propagation axis and is associated with a prominent spatially dependent phase structure \cite{allen1992orbital}. 

In the next step, we discretize the continuous spatial space of signal and idler in Eq. \eqref{GeneralBiphotonstate} with LG modes to investigate the OAM mode distribution of the biphoton state. Mathematically, this corresponds to the basis expansion
\begin{equation}
    \ket{\psi} = \sum_{p_{\signal},\ell_{\signal}}  \sum_{p_{\idler},\ell_{\idler}} C_{p_{\signal},p_{\idler}}^{\ell_{\signal},\ell_{\idler}} \biggl( \ket{p_{\signal},\ell_{\signal}} \otimes \ket{p_{\idler},\ell_{\idler}} \biggl),
    \label{LGdecomposition}
\end{equation}
where $\ket{p,\ell} = \int d\bm{q} \:  \text{LG}_{p}^{\ell}(\bm{q},w) \: \hat{a}^{\dagger}(\bm{q}) \ket{\text{vac}}$ is the single-photon state representation of a LG mode. The expansion coefficients of the LG basis $C_{p_{\signal},p_{\idler}}^{\ell_{\signal},\ell_{\idler}}$ for an LG pump beam are given by \cite{baghdasaryan2022generalized, yao2011angular}
\begin{eqnarray}
    C^{\ell_{\signal},\ell_{\idler}}_{p_{\signal},p_{\idler}} && \bigl( \underbrace{p_{\pump},\ell_{\pump},w_{\pump}}_{\text{pump}}; \underbrace{\Phi_{\text{PM}},L}_{\text{crystal}}; \underbrace{w_{\signal},w_{\idler}}_{\text{collection}} \bigl)  = \biggl( \bra{p_{\signal},\ell_{\signal}} \otimes \bra{p_{\idler},\ell_{\idler}} \biggl) \ket{\psi} \nonumber \\
    && =\mathcal{N}  \iint d\bm{q}_{\signal} \: d\bm{q}_{\idler} \: \text{LG}_{p_{\mathrm{p}}}^{\ell_{\pump}}(\bm{q}_{\signal}+\bm{q}_{\idler}, w_{\pump}) \: \Phi_{\text{PM}}(\Delta k_z) \nonumber \\ 
    && \hspace{1.0cm} \times \left[ \text{LG}_{p_{\signal}}^{\ell_{\signal}}(\bm{q}_{\signal},w_{\signal}) \right]^* \left[ \text{LG}_{p_{\idler}}^{\ell_{\idler}} (\bm{q}_{\idler},w_{\idler}) \right]^*.
\label{OverlapAmplitudes}
\end{eqnarray}
The joint probability of a projection into the LG modes $\ket{p_{\signal},\ell_{\mathrm{s}}}$ for signal and $\ket{p_{\idler},\ell_{\mathrm{i}}}$ for idler is $P_{p_{\signal},p_{\idler}}^{\ell_{\signal},\ell_{\idler}} = |C_{p_{\signal},p_{\idler}}^{\ell_{\signal},\ell_{\idler}}|^2$. 

In addition to the pump beam and phase-matching inside the crystal, the expansion amplitude depends on the collection optics used in the projection scheme \cite{bernecker2023spatial}. The spatial state of the signal and idler is often examined trough mode converters \cite{agnew2011tomography,sevilla2024spectral}. 

For instance, SLMs display computer-generated holograms that convert an incident LG mode, based on the encoded phase profile, into the fundamental Gaussian mode (FGM), $p=\ell=0$. The FGM can be efficiently coupled into a single-mode fiber (SMF) and registered at a photodetector. If other orthogonal LG modes are incident on the SLM, the same hologram produces a light mode orthogonal to the FGM, preventing a coupling into the SMF. Through this measurement procedure, the collected signal and idler modes possess defined beam sizes $w_{\signal}$ and $w_{\idler}$.

The accurate measurement of the radial part of LG modes was demonstrated to be experimentally challenging since the transverse structure and the orthogonality necessitate an amplitude-sensitive detection. LG modes with radial indices $p>0$ are difficult to be projected efficiently since their corresponding holograms have to modulate a more sophisticated phase and amplitude structure compared to the case $p=0$. \cite{salakhutdinov2012full, valencia2021entangled}. Methods like multi-plane-light-conversion (MPLC) \cite{fontaine2019laguerre, hiekkamaki2019near, sevilla2024spectral} have achieved much progress in the efficient detection of full-field modes, but for experimental feasibility most set-ups concentrate on the OAM part for spatial entanglement and collect only modes with $p_{\signal}=p_{\idler}=0$.  This corresponds to the projection 
\begin{eqnarray}
     \ket{\psi ^{\prime}} &&=\biggl(\ket{0,\ell_{\signal}} \otimes \ket{0,\ell_{\idler}} \biggl) \biggl( \bra{0,\ell_{\signal}} \otimes \bra{0, \ell_{\idler}} \biggl) \ket{\psi} \nonumber \\
     &&= \sum_{\ell_{\signal},\ell_{\idler}} C^{\ell_{\signal},\ell_{\idler}}_{0,0} \ket{\ell_{\signal}} \otimes \ket{\ell_{\idler}} :=\sum_{\ell_{\signal},\ell_{\idler}} C^{\ell_{\signal},\ell_{\idler}} \ket{\ell_{\signal}, \ell_{\idler}}. 
     \label{StateProjection}
\end{eqnarray}
We abbreviate $\ket{\ell_{\signal}} \otimes \ket{\ell_{\idler}}:=\ket{\ell_{\signal},\ell_{\idler}}$ as the biphoton OAM mode, meaning signal and idler are in the radial projected LG modes $\ket{\ell_{\signal}}:= \ket{0,\ell_{\signal}}$ and $\ket{\ell_{\idler}}:= \ket{0,\ell_{\idler}}$. Thus, we define the joint collection probability of the biphoton OAM mode as $P^{\ell_{\signal},\ell_{\idler}}= \Bigl| C^{\ell_{\signal},\ell_{\idler}} \Bigl|^2$. The probability distribution in terms of the discrete integers $\ell_{\signal}$ and $\ell_{\idler}$ describes the OAM spectrum. Theoretical and experimental studies \cite{mair2001entanglement, walborn2004entanglement} showed, that the OAM conservation law $\ell_{\pump}=\ell_{\signal}+\ell_{\idler}$ is fulfilled in quasi-collinear regimes. All generated biphoton OAM modes $\ket{\ell_{\signal},\ell_{\idler}}$ obey this conservation rule but have in general different probabilities $P^{\ell_{\signal},\ell_{\idler}}$.

\begin{figure*}
\includegraphics[width=\linewidth]{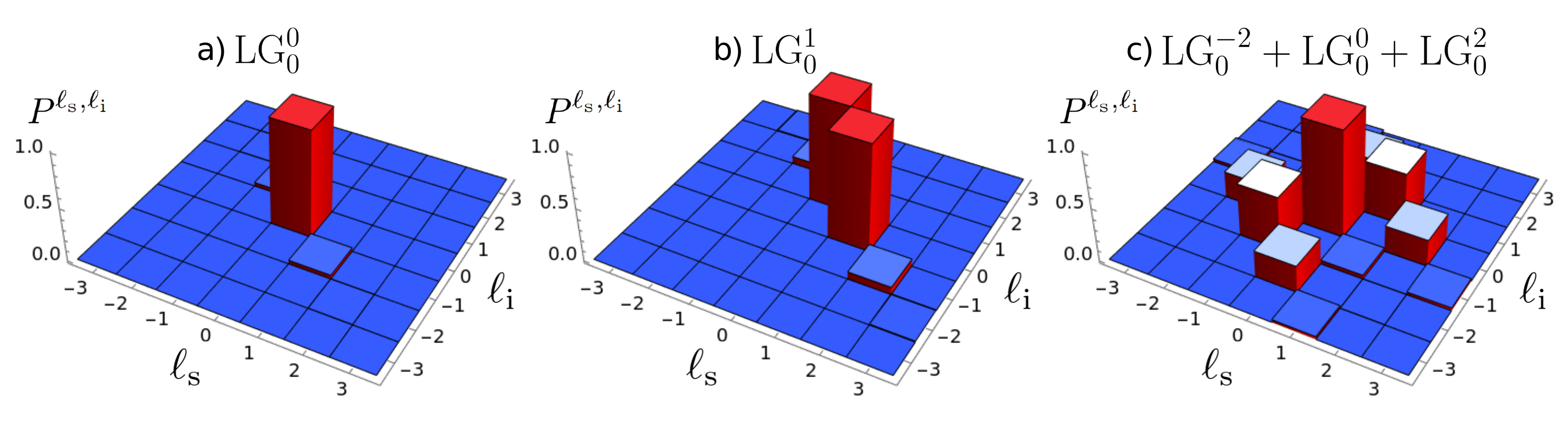}
\caption{\label{figure1} 
Normalized OAM spectra within the range $\ell_{\signal}, \ell_{\idler}=-3, ..., 3$ for different pump beams. a) \& b) Depending on the pump OAM index of the $\text{LG}_0^{\ell_{\pump}}$ mode, biphoton OAM modes $\ket{\ell_{\signal},\ell_{\idler}}$ with different probabilities $P^{\ell_{\signal}, \ell_{\idler}}$ are generated along a single anti-diagonal obeying $\ell_{\pump}=\ell_{\signal}+\ell_{\idler}$. c) A superposition of multiple LG modes nests several anti-diagonals in the OAM spectrum. Here, we show an equal-weighted superposition of LG modes with $\ell_{\pump}=-2,0,2$.}
\end{figure*}

\subsection{\label{subsec:Maximal entanglement}Maximal entanglement}

A pure quantum state of a bipartite system that cannot be factored as a tensor product of states of its local constituents is called entangled. It has to be considered as an inseparable whole. 

A particularly convenient measure of entanglement is the Schmidt number $K$ obtained by the composition of a bipartite state into its Schmidt form \cite{miatto2012cartesian}
\begin{equation}
    \ket{\Psi} = \sum_{k=1}^r \lambda_k \ket{\signal_k} \otimes \ket{\idler_k}. 
\label{SchmidtDecomposition}
\end{equation}
Here, $r$ is the Schmidt rank, $\lambda_k$ are the Schmidt coefficients, and $\ket{\signal_k}, \ket{\idler_k} $ are the Schmidt modes for the subsystems signal and idler, respectively. The normalization condition is $ \sum_k \lambda_k^2 =1$. The Schmidt decomposition represents the joint basis that maximizes the mutual dependence of two variables in the two subsystems \cite{miatto2012spatial}. Because of the single sum, any entangled state is characterized by a rank $r>1$ and allows the definition of the Schmidt number $K = 1/ \sum_k \lambda_k^4$. A normalized state is fully separable for $K=1$ and indicates the absence of entanglement. The more the Schmidt coefficients are distributed in the sum of Eq. \eqref{SchmidtDecomposition}, the more entangled the state is, and $K$ increases. 

The Schmidt number for a state of two qudits ($\mathcal{H}=\mathbb{C}^d$) has an upper bound of $K=d$, implying maximal entanglement. Maximally entangled states (MESs) have a maximal rank $r=d$ and are characterized by equal singular values $\lambda_k=1/\sqrt{d}$, leading to complete non-separability. In the OAM basis, MESs are represented by a matrix in Eq. \eqref{StateProjection} that is proportional to a permutation matrix, where each column and row contains a single entry of $1/\sqrt{d}$.

\subsection{Subspaces and entanglement filtering}

Limiting the sum in Eq. \eqref{StateProjection} to a finite cut-off provides a practical setting for the experimental implementation of OAM qudits. Most applications in quantum communication utilize measurements that are sensitive only to photons prepared in a specific subspace (e.g. with probabilistic mode analyzers \cite{groblacher2006experimental} or filter measurements \cite{mafu2013higher}). When the corresponding holograms of OAM modes (or their superpositions) within a specific subspace are displayed on SLMs and used for state projection or quantum state tomography, only these modes are effectively registered in the counting statistics. All modes outside of a subspace do not contribute to the measured outcomes. Their overlap with all used holograms (in the specified basis) is certain to be not coupled into the SMF due to the completeness and orthogonality.

In this analysis, we restrict ourselves to equal subspaces for signal and idler, i.e. we investigate the single-photon Hilbert spaces $\mathcal{H}_{\signal}=\mathcal{H}_{\idler}=\mathbb{C}^d$ of dimension $d$. The total bipartite system is described in a two-photon Hilbert space formed by the tensor product $\mathcal{H}_{\signal} \otimes \mathcal{H}_{\idler}$ with dimensionality $d^2$. 

A well-known approach to enhance the entanglement between signal and idler is to create equal probable OAM modes with entanglement concentration, also known as the Procrustean method. Say we aim to maximize the entanglement of a state described by non-equal, but also non-vanishing expansion amplitudes from Eq. \eqref{StateProjection}:
\begin{equation}
    C^1, C^2, \dots, C^n,
    \label{amplitudeswithout}
\end{equation}
where $1,...,n$ labels all possible ordered pairs of OAM indices $\ell_{\signal}$ and $\ell_{\idler}$ of interest. 

The Procrustean method is a filtering technique that equalizes 
these mode amplitudes with local operations behind the crystal to either one of the signal or idler beam paths, or both beam paths by altering the measurement efficiency with optical components. These local operations transform the amplitudes \eqref{amplitudeswithout} from above into the modified amplitudes \cite{chen2020coherent}
\begin{equation}
    x_1 \cdot C^1, x_2 \cdot C^2, \dots,  x_n \cdot C^n
\end{equation}
where $x_1,..., x_n$ depends on the exact set-up and can be chosen to equalize $x_1 \cdot C^1 = x_2 \cdot C^2 = \dots = x_n \cdot C^n$. Experimentally, lenses \cite{vaziri2003concentration} or specially designed holographic gratings \cite{dada2011experimental, chen2020coherent} that adjust the diffraction efficiencies have been used to demonstrate entanglement concentration. The disadvantage of this amplitude filtering is an enormous additional effort in set-up and a significant reduction of the detection rate since Procrustean filtering introduces mode selective loss. These states generated by post-filtering are not indeed MESs \cite{liu2018coherent}. 

On the contrary, we investigate how to shape the spatial profile of the pump and the phase-matching function of the nonlinear structures can tune OAM amplitudes directly. 
These are \textit{genuine} MESs without the need for external optical elements to manipulate the biphoton state.

\section{\label{sec:Engineering methods}Quantum state engineering }

We conduct our proof-of-principle investigation for degenerate, collinear type II SPDC in the narrowband regime with $k_{\pump}=2k_{\signal}=2k_{\idler}$. In this setup, a $\lambda_{\pump}= \SI{405}{\nano \meter}$ pump with a beam waist $w_{\pump}= \SI{25}{\micro \meter}$ produces photon pairs in a $L= \SI{15}{\milli  \meter}$ long KTP crystal. The signal and idler photons are assumed to be collected in modes with sizes $w_{\signal}=w_{\idler}= \SI{33}{\micro \meter}$. The experimental parameters are adapted from Ref. \cite{kovlakov2018quantum}.

\subsection{\label{subsec:Pump Engineering}Pump Engineering}

The ability to experimentally modulate the amplitude and phase of spatial light modes offered new perspectives to engineer an arbitrary superposition of Laguerre-Gauss (LG) modes as a pumping source for SPDC. Each LG mode in this superposition can be manipulated independently, and based on OAM conservation $\ell_{\pump}=\ell_{\signal}+\ell_{\idler}$, we can predict which signal and idler modes are generated. All modes of non-zero detection probability for a single pump OAM index are positioned along anti-diagonal lines, like shown in Fig. \ref{figure1} for $\ell_{\pump}=0$ in a) and for $\ell_{\pump}=1$ in b). When considering a pump superposition of LG modes, the OAM spectrum is nested with modes along \textit{several} anti-diagonal lines.

Fig. \ref{figure1} c) displays a typical visualization of the OAM spectrum for an equal-weighted pump superposition of radial-projected LG modes with $\ell_{\pump}=-2,0,2$, where only the most significant contributions of OAM biphoton modes $\ket{\ell_{\signal}, \ell_{\idler}}$ in a range of  $\ell_{\signal}, \ell_{\idler}=-3,...,3$ are shown. Note that setting $k_{\pump}=2k_{\signal}=2k_{\idler}$ introduces interchangeability of the OAM indices for the expansion amplitudes $C^{\ell_{\signal}, \ell_{\idler}} = C^{\ell_{\idler}, \ell_{\signal}}$,which leads to equal probabilities  $P^{\ell_{\signal}, \ell_{\idler}} = P^{\ell_{\idler}, \ell_{\signal}} $ and a symmetric OAM spectrum with respect to $\ell_{\signal}$ and $\ell_{\idler}$.

Experimentally, we are not restricted to equal weightings, as we can encode corresponding holograms for any superposition on the SLM. Thus, we introduce free pump coefficients $a_{\ell_{\pump}} \in \mathbb{C}$ into the updated angular spectrum from Eq. \eqref{PumpinLG}:
\begin{equation}
    V_{\pump}(\bm{q}_{\pump}) = \sum_{\ell_{\pump}} a_{\ell_{\pump}}  \: \text{LG}_0^{\ell_{\pump}}(\bm{q}_{\pump},w_{\pump}).
    \label{PumpinLGsuperposition}
\end{equation}
The OAM spectrum can be reconfigured by adjusting the values of $a_{\ell_{\pump}}$. Suitable pump coefficients for OAM modes of equal probability were determined using simultaneous perturbation stochastic approximation algorithms \cite{kovlakov2018quantum}. In addition, Ref. \cite{baghdasaryan2022generalized} showed that with the expression for $C^{\ell_{\signal},\ell_{\idler}}$, the pump coefficients can be calculates analytically.

By inserting the LG superposition \eqref{PumpinLGsuperposition} into Eq. \eqref{OverlapAmplitudes} and \eqref{StateProjection}, we can write the state because of OAM conservation as
\begin{eqnarray}
    \ket{\psi^{\prime}} &&= \sum_{\ell_{\pump},\ell_{\signal},\ell_{\idler}}^{} \: a_{\ell_{\pump}} \: \delta_{\ell_{\pump},\ell_{\signal}+\ell_{\idler}} \: C^{\ell_{\signal},\ell_{\idler}} \ket{\ell_{\signal}, \ell_{\idler}} \nonumber \\
    &&=\sum_{\ell_{\signal},\ell_{\idler}}^{} \: a_{\ell_{\signal}+\ell_{\idler}} \: C^{\ell_{\signal},\ell_{\idler}} \ket{\ell_{\signal}, \ell_{\idler}}.
    \label{GeneralLGsuperposition}
\end{eqnarray}
As an example, we study the generation of the MES $\ket{\Psi_1} = \frac{1}{\sqrt{3}} ( \ket{-1,-1} + \ket{0,0} + \ket{1,1} )$ like in Refs. \cite{kovlakov2018quantum, liu2018coherent}. We can represent the modified amplitudes of the state \eqref{GeneralLGsuperposition} in a matrix form with matrix elements $A_{\ell_s,\ell_i}= a_{\ell_{\signal}+\ell_{\idler}} \: C^{\ell_{\signal},\ell_{\idler}}$ and compare it with a similar matrix for the state  $\ket{\Psi_1}$. If we restrict the summations in Eq. \eqref{GeneralLGsuperposition} to OAM indices $\ell_{\signal}, \ell_{\idler}=-1,0,1$ , we end up with matrices:
\begin{equation*}
\begin{pmatrix}
a_{-2} \: C^{-1,-1} & a_{-1}  \: C^{0,-1} & a_{0} \: C^{1,-1}  \\
a_{-1}  \: C^{-1,0} & a_{0}  \: C^{0,0} & a_{1} \: C^{1,0}   \\
a_{0}  \: C^{-1,1} & a_{1}  \: C^{0,1} & a_{2} \: C^{1,1}  \\
\end{pmatrix} 
\overset{!}{=} \frac{1}{\sqrt{3}} \begin{pmatrix}
1 & 0 & 0 \\
0 & 1 & 0  \\
0 & 0 & 1 \\
\end{pmatrix}. \\
\end{equation*}
The comparison of coefficients directly indicates $a_{-1}=a_{1}=0$ as we clearly do not intend to generate modes with $\ell_{\signal}+\ell_{\idler}=\pm 1$.  Furthermore, it follows $a_{-2} \propto 1/C_{-1,-1}, a_{0}\propto1/C_{0,0}$, and $ a_{2}\propto 1/ C_{1,1}$.

\begin{figure}[t]
\includegraphics[width=0.8\linewidth]{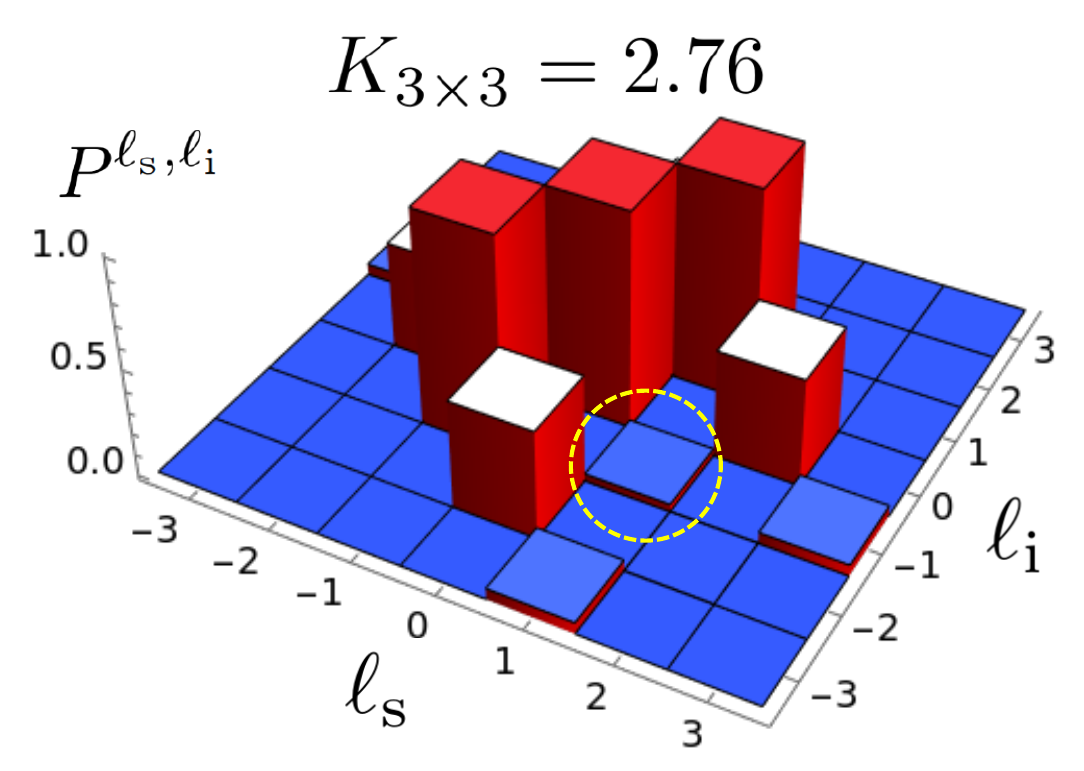}
\caption{\label{figure2} 
The pump-engineered OAM spectrum from Fig. \ref{figure1} c) within the range $\ell_{\signal}, \ell_{\idler}=-3, ..., 3$. The target state $\ket{\Psi_1}=\frac{1}{\sqrt{3}} \bigl( \ket{-1,-1}+\ket{0,0}+\ket{1,1} \bigl)$ is not maximally entangled in the subspace $S_{3 \times 3}$ (corresponding to $\ell_{\signal}=\ell_{\idler}=-1,0,1$), due to the presence of the unintended modes $\ket{1,-1}$ (yellow circle) and $\ket{-1,1}$ (not visible). The Schmidt number is $K_{3 \times 3}=2.76<3$.}
\end{figure}

The OAM spectrum with the pump coefficients determined above can be seen in Fig. \ref{figure2}. The biphoton OAM modes $\{\ket{-1,-1}, \ket{0,0}, \ket{1,1} \}$ have equal expansion amplitudes and projection probabilities, as desired. However, this condition is not sufficient for maximal entanglement. Since $C_{-1,1} = C_{1,-1} \neq 0$ and $a_0  \neq 0$, the modes $\ket{-1,1}$ and $\ket{1,-1}$ will have non-vanishing probabilities leading to a decrease of entanglement. In order to analyze the entanglement, we consider the Hilbert space $\mathcal{H}_{\signal} = \mathcal{H}_{\idler} = \{\ket{-1}, \ket{0}, \ket{1} \}$, resulting in a nine-dimensional two-photon subspace 
\begin{eqnarray*}
    \mathcal{H}_{\signal} \otimes \mathcal{H}_{\idler} = && \Bigl \{ \ket{-1,-1}, \ket{-1,0}, \ket{-1,1}, \ket{0,-1}, \\
    && \ket{0,0},\ket{0,1}, \ket{1,-1},\ket{1,0}, \ket{1,1} \Bigl \}.
\end{eqnarray*}
We denote this subspace as $S_{3 \times 3}$ because each subsystem has three levels. 

The OAM spectrum in Fig. \ref{figure2} analyzed in this subspace reveals two types of modes. On the one hand, we have \textit{target modes}, which have equal probability and contribute to desired MESs like $\ket{-1,-1}, \ket{0,0}, \ket{1,1}$ in our example. On the other hand, we find \textit{unintended modes}, which are still present despite pump engineering due to OAM conservation, like $\ket{-1,1}$ and $\ket{1,-1}$. If we calculate the Schmidt number in $S_{3 \times 3}$, denoted by $K_{3 \times 3}$, we get $K_{3 \times 3}=2.76$ due to non-vanishing contributions of the unintended modes. An actual MES would require $K_{3 \times 3}=3$. The suppression of the unwanted expansion amplitudes $C^{-1,1}$ and $C^{1,-1}$ is not possible with further pump engineering.

\begin{figure}[t]
\includegraphics[width=0.65\linewidth]{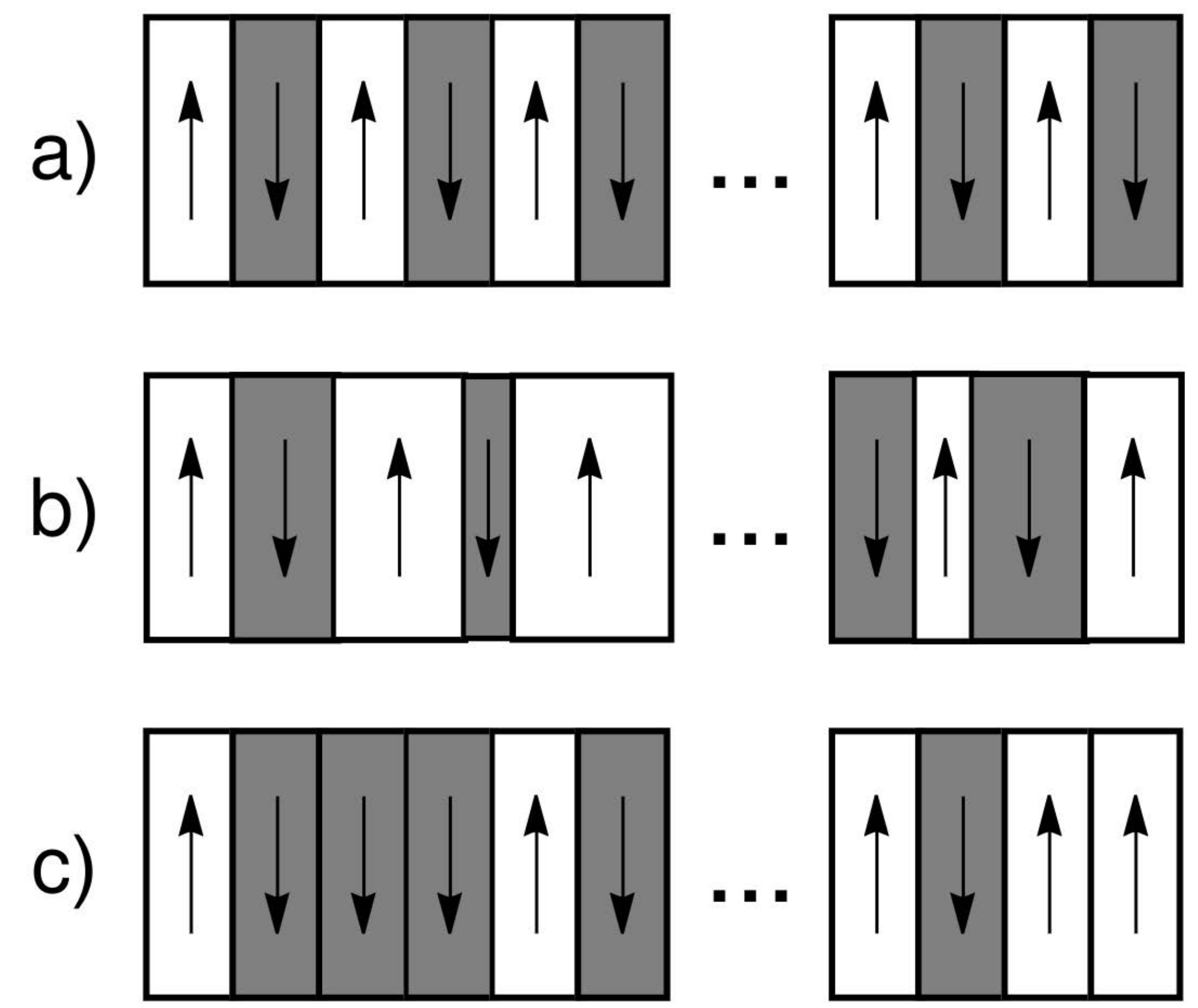}
\caption{\label{figure3}
Conceptual sketch of different crystal-domain configurations to achieve nontrivial quasi-phase-matching. a) Periodic poling: domains of equal width are poled in an alternating pattern of a positive ($\uparrow$) or negative ($\downarrow$) effective nonlinear susceptibility. b) Domain-width annealing: use of a periodically alternating pattern, but each domain width is individually adjusted.  c) Costume poling: the domain width is kept fixed, but the orientation pattern is customized as desired.}
\end{figure}

\subsection{\label{subsec:Tailored nonlinear crystals}Tailored nonlinear crystals}
\subsubsection{Enginnering the phase-matching function}
The idea of tailoring the nonlinearity of crystals used in SPDC caught much attention lately due to novel fabrication methods \cite{graffitti2017pure}. Instead of a single nonlinear material with susceptibility $\chi^{(2)}$, we consider a medium decomposed in thin layers of the birefringent material, so-called domains. Each domain is poled with an either up ($+\chi_{\text{eff}}$) or down ($-\chi_{\text{eff}}$) oriented effective susceptibility, so the nonlinearity profile along the propagation axis of the beams $\chi^{(2)}(z)$ results in a discontinuous function. However, the corresponding PMF for the entire crystal is the linear superposition of the PMF of each individual domain \cite{dosseva2016shaping}.

\begin{figure*}
\includegraphics[width=\linewidth]{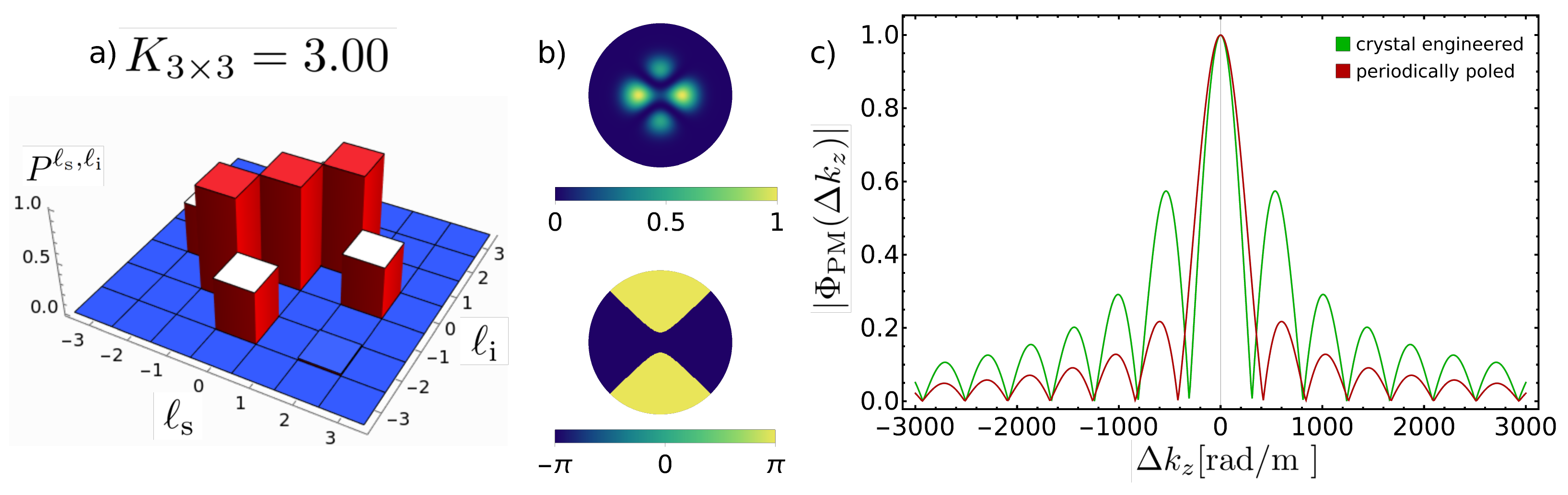}
\caption{
\label{figure4} 
a) The normalized OAM spectrum for the target state $\ket{\Psi_1}=\frac{1}{\sqrt{3}} \bigl( \ket{-1,-1}+\ket{0,0}+\ket{1,1} \bigl)$ within the range $\ell_{\signal}, \ell_{\idler}=-3, ..., 3$. The state is maximally entangled in the subspace $S_{3 \times 3}$ by combined pump \textit{and} crystal engineering. The unintended modes $\ket{1,-1}$ and $\ket{-1,1}$ as observed in Fig. \ref{figure2} are suppressed. b) The corresponding intensity and phase profile of the LG superposition with pump coefficients $a_{-2}=a_{2}=2.46$ and $a_{0}=1.73$. c) The customized PMF $\Phi_{\text{PM}}(\Delta k)$ with crystal coefficients $c_0=1.045$ and $c_1=-0.865$ (green) and for comparison, a typical sinc-like PMF of a periodically poled crystal (red).
}
\end{figure*}

Periodically poled crystals, with alternating domains of uniform length as in Fig. \ref{figure3} a), emerged as highly effective in maintaining quasi-phase-matching (QPM) for a wide range of wavelengths in different nonlinear materials \cite{fejer1992quasi, fradkin1999tunable}. This configuration results in an approximate sinc-like PMF. Since periodically poled crystals are designed to optimize the brightness of photon pair sources, alternative  PMFs realized with the crystal would automatically generate fewer pairs with the same pump beam. 

However, the implementation of different PMFs has been investigated lately and tested experimentally. The PMF can be shaped using \textit{nontrivial} QPM, where the constant poling pattern is maintained, but the domain width is individually adapted \cite{dixon2013spectral, cui2019wave} (so-called domain width-anneling, see Fig. \ref{figure3} b)) or different domains exhibit different values for $\chi_{\text{eff}}$ \cite{branczyk2011engineered}. Alternatively, each domain orientation can be regulated individually \cite{tambasco2016domain} so that the effective PMF can match nearly any desired target function by customizing the poling configuration \cite{dosseva2016shaping} like in Fig. \ref{figure3} c). All approaches successfully demonstrated high-purity single-photon generation via SPDC without post-filtering due to minimized spectral \cite{graffitti2018independent, pickston2021optimised} or spatial \cite{baghdasaryan2023enhancing} correlations between signal and idler photons. Custom poled structures in waveguides like the multi-output quantum pulse gate (MGPQ) \cite{serino2023realization} also proved beneficial for decoding high-dimensional single-photon temporal modes. While higher pump powers can adjust the reduced MQPG efficiency \cite{brecht2014demonstration}, this may amplify multi-photon effects \cite{takeoka2015full}.

Here, we leverage the ability to tailor PMFs. Similar to the pump beam construction from the previous section, we assume a general function for susceptibility function $\chi^{(2)}(z)$ by expanding it into a cosine Fourier series like in Ref. \cite{baghdasaryan2023enhancing}:
\begin{equation}
    \chi^{(2)}(z) = \sum_{n=0}^N c_n \cos{ \Bigl(\frac{n z}{\sigma}  \Bigl)},
    \label{CosineSus}
\end{equation}
where the \textit{crystal coefficients }$c_n$ are undetermined. The scaling factor $\sigma=L/4$ insures high single-photon purity without significant reduction in the source brightness \cite{tambasco2016domain, graffitti2017pure, baghdasaryan2023enhancing}. In analog to pump engineering, we treat $c_n$ as free parameters and \textit{tune} them to equalize or suppress expansion amplitudes in Eq. \eqref{OverlapAmplitudes}. The according PMF when inserting Eq. \eqref{CosineSus} in Eq. \eqref{Phase-matching-Function} reads
\begin{eqnarray}
    \Phi_{\text{PM}} (\Delta k) 
    = \frac{L}{2} \: \sum_{n=0}^N && c_n \Biggl( \: \text{sinc} \Bigl( \frac{\Delta k L}{2} + 2n \Bigl) \nonumber \\
    && + \: \text{sinc} \Bigl( \frac{\Delta k L}{2 } -2n \Bigl) \Biggl),
\label{CosinePMF}
\end{eqnarray}
so we fully parameterize the PMF by the crystal coefficients $c_n$. The familiar sinc-PMF is regained for $c_{0}=1$ while $c_{n>0}=0$.

\subsubsection{Constraints by the relative mode number}
We assume, for simplicity, that  $w_{\pump} \approx \sqrt{\frac{L}{k_{\pump}}}$ and $\sqrt{2} w_{\pump} \approx w_{\signal}=w_{\idler}$, which can equivalently be enforced by assuming degenerate SPDC $k_{\mathrm{p}}=2k_{\mathrm{s}}=2k_{\mathrm{i}}$ and demanding equal Rayleigh lengths for pump, signal, and idler. Experimentally, this setting is favored to concentrate the spread of OAM modes along the anti-diagonal $\ell_{\pump}=0$ to the mode $\ket{0,0}$ \cite{kovlakov2018quantum}.
Under this consideration, we find a explicit expression for Eq. \eqref{CosineSus} from Ref. \cite{baghdasaryan2022generalized} as
\begin{small}
\begin{equation}
    C_{}^{\ell_{\signal},\ell_{\idler}}(N_R) \propto \int_{-\frac{L}{2}}^{\frac{L}{2}} \: dz \: c_n \: \underbrace{\frac{(k_{\pump} w_{\pump}^2 + 2iz)^{N_R}}{(k_{\pump} w_{\pump}^2 - 2iz)^{N_R+1}}}_{:= \xi(N_R,z)} \cos{ \Bigl(\frac{n z}{\sigma}  \Bigl)} ,
    \label{RMNinExpansionAmplitude}
\end{equation}
\end{small}
where the relative mode number (RMN) $N_R$  appears in the $z$-dependent function $\xi(N_R,z)$. 

The RMN was already discussed in the context of decoupling the spatial and spectral DOFs in the SPDC process \cite{baghdasaryan2022generalized}. In our investigation the value is $N_R = |\ell_{\signal}+\ell_{\idler}|-|\ell_{\signal}|-|\ell_{\idler}|$.

Eq. \eqref{RMNinExpansionAmplitude} implies that the expansion amplitudes of biphoton OAM modes are proportional when their RMNs are equal. This means that OAM modes with the same RMN cannot be manipulated independently, which limits our state engineering capabilities. Specifically, if $N_R = N_R^{\prime}$, we cannot find a set of crystal coefficients $c_n$ to suppress $C_{}^{\ell_{\signal},\ell_{\idler}}(N_R)$ while ensuring $C_{}^{\ell_{\signal}^{\prime},\ell_{\idler}^{\prime}}(N_R^{\prime}) \neq 0.$ 

Consider for example the generation of the target state $\ket{\Psi_2}=\frac{1}{\sqrt{3}} \bigl( \ket{-1,0} + \ket{0,1} + \ket{1,-1} \bigl)$, which is a MES in $S_{3 \times 3}$. The existence of the desired mode $\ket{1,-1}$ governs the pump superposition to have a pump coefficient $a_0 \neq 0$, which in turn means the unintended mode $\ket{0,0}$ appears in the OAM spectrum. Eq. \eqref{RMNinExpansionAmplitude} suggests that if we can find crystal coefficients $c_n$ to suppress  $C^{0,0} \approx 0$, we will also suppress the target mode $C^{0,1} \approx 0$, as both modes fulfill $N_R=0$.

Therefore, crystal engineering can optimize the generation of MESs when all unintended modes in the considered subspace have a different RMN than the target modes. More details on RMN and general guidelines for selecting MESs that follow this rule can be found in the Appendix \ref{appendixRMN}.

\section{\label{subsec:Results}Results}

\begin{figure*}
\includegraphics[width=\linewidth]{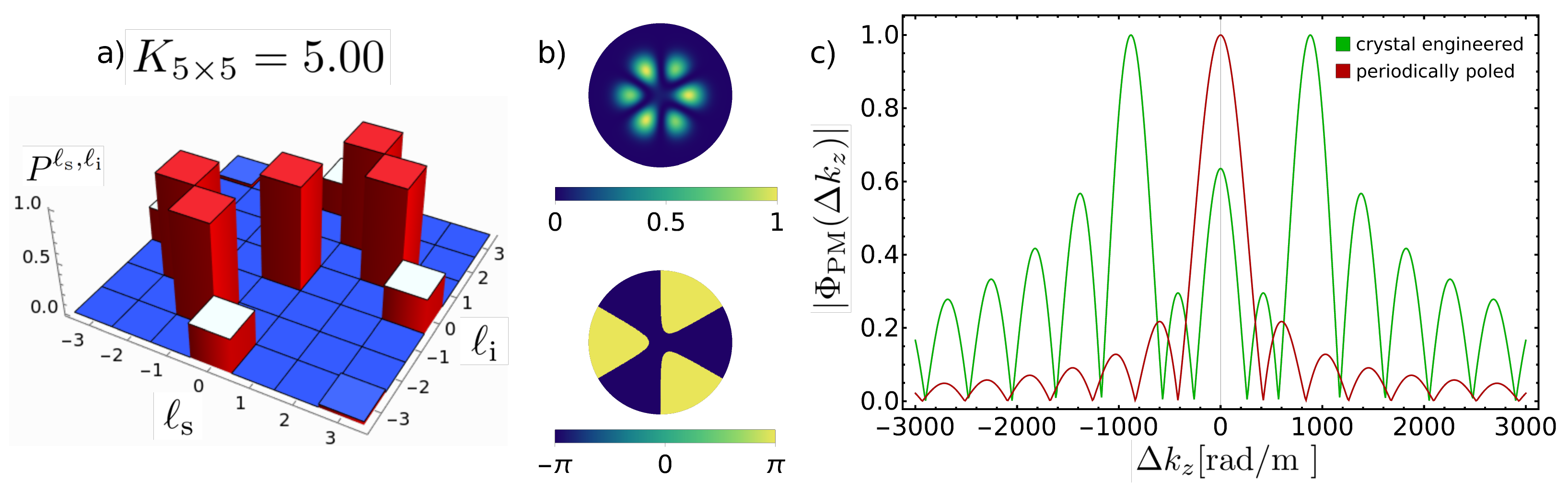}
\caption{\label{figure5} 
Example for a maximally entangled two-ququint state ($d=5$).
a) The normalized OAM spectrum for the target state $\ket{\Psi_3}= \frac{1}{\sqrt{5}} \bigl( \ket{-2,-1} + \ket{-1,-2} + \ket{0,0} + \ket{1,2} + \ket{2,1} \bigl)$ within the range $\ell_{\signal}, \ell_{\idler}=-3, ..., 3$. The state is maximally entangled in the subspace $S_{5 \times 5}$. b) The required pump superposition is given by the pump coefficients $a_{-3}=a_3=3.68$ and $a_0=2.24$. c) The corresponding PMF $\Phi_{\text{PM}}(\Delta k)$ with $c_0=1.604, c_1=-1.665$, and $c_2=0.912$ (green) and for comparison, a typical sinc-like PMF of a periodically poled crystal (red). Note that here more crystal coefficients are needed to find the optimal PMF than in Fig. \ref{figure4}.
}
\end{figure*}

\subsection{Combined pump and crystal engineering}

As we have discussed before, the target state $\ket{\Psi_1}=\frac{1}{\sqrt{3}} \bigl( \ket{-1,-1}+\ket{0,0}+\ket{1,1} \bigl)$ is not maximally entangled in the subspace $S_{3 \times 3}$ (see Fig. \ref{figure2}) if we only apply pump engineering.
Since all three target modes have the RMN $N_R=0$ and the unintended modes $\ket{-1,1}$ und $\ket{1,-1}$ have $N_R=-2$, we can utilize crystal engineering to suppress $\ket{-1,1}$ und $\ket{1,-1}$. Thus we like to engineer $C^{-1,1}=C^{1,-1}=0$ while $C^{-1,-1}=C^{0,0}=C^{1,1}=1/\sqrt{3}$. The latter condition can be reached by pure pump engineering. The first condition results in a linear system of equations obtained by Eq. \eqref{RMNinExpansionAmplitude} (see Appendix \ref{appendixRMN}):
\begin{eqnarray}
&&\frac{1}{\sqrt{3}} 
\begin{pmatrix}
1 \\
0 \\
\end{pmatrix} \overset{!}{=} 
\begin{pmatrix}
C^{0,0}(N_R=0) \\
C^{-1,1}(N_R=-2) \\
\end{pmatrix} 
\nonumber
\propto
\\[0.15cm] 
&& 
\begin{pmatrix}
\int dz \: \xi(0,z)  & 
\int dz \:\xi(0,z) \cos{(\frac{z}{\sigma})}  \\
\int dz \:\xi(-2,z) & 
\int dz \: \xi(-2,z)\cos{(\frac{z}{\sigma})}  
\end{pmatrix} 
\cdot
\begin{pmatrix}
c_0 \\
c_1 
\end{pmatrix} 
\label{matrixforstate3.1}
\end{eqnarray}
with solution $c_0=1.045$ and $c_1=-0.865$. The  condition $C^{1,-1}=C^{-1,1}$ is automatically fulfilled due to $k_{\signal}=k_{\idler}$. The modified OAM spectrum by the adapted PMF using the calculated crystal coefficients is displayed in Fig. \ref{figure4} a). The engineered state is now a MES with Schmidt number $K_{3 \times 3}=3$.

Additionally, Fig. \ref{figure4} plots the corresponding spatial distribution of the pump beam and the normalized PMF in b) and c), respectively.  The pump configuration remains unchanged as determined in Section \ref{subsec:Pump Engineering}. The four intensity clumps are in agreement with experimental demonstrations \cite{kovlakov2018quantum, liu2018coherent}. The equal weights $a_{-2}=a_2$ in the LG superposition can be explained by the property $C^{\ell_{\signal},\ell_{\idler}}=(C^{-\ell_{\signal},-\ell_{\idler}})^*$ \cite{baghdasaryan2022generalized}: the anti-diagonals $\ell_{\pump}=-2$ and $\ell_{\pump}=2$ are adjusted  similarly to equalize $C^{-1,-1}$ and $C^{1,1}$. Therefore, the superimposed phases of $\text{LG}_0^{-2}$ and $\text{LG}_0^{2}$ results in a phase profile of $\pi$ and $-\pi$ only. The calculated crystal coefficients $c_0$ and $c_1$ suggest that the best choice to generate the target state differs from a periodically poled crystal profile. For comparison, we show both PMFs for a crystal of length $L= \SI{15}{\milli  \meter}$. The customized function for maximal entanglement (green) has more pronounced outer sidelobes compared to the sinc-function of a periodically poled crystal (red).

Besides the MES $\ket{\Psi_1}=\frac{1}{\sqrt{3}} \bigl( \ket{-1,-1}+\ket{0,0}+\ket{1,1} \bigl)$, an infinite number of states can be inferred which fulfill $K_{3 \times 3}=3$ when taking the relative phases between the modes into account: $\ket{\Psi_1^{\prime}}=\frac{1}{\sqrt{3}} \bigl( \ket{-1,-1}+ \mathrm{e}^{i \kappa_1} \ket{0,0}+ \mathrm{e}^{i \kappa_2} \ket{1,1} \bigl)$ with arbitrary $\kappa_1, \kappa_2 \in \mathbb{R}$.
Such relative phases can be inherited by the LG superposition of the pump beam \cite{liu2018coherent} due to the linearity in Eq. \eqref{GeneralLGsuperposition}, where all biphoton OAM modes $\ket{\ell_{\signal},\ell_{\idler}}$ with $\ell_{\pump}=\ell_{\signal}+\ell_{\idler}$ obtain the same prefactor from $a_{\ell_{\pump}}$.
We can introduce arbitrary $\mathrm{e}^{i \kappa_{1,2}}$ for the mode $\ket{\ell_{\signal},\ell_{\idler}}$ by crystal engineering as well, but the mode $\ket{\ell_{\idler},\ell_{\signal}}$ simultaneously acquire the same prefactor.

\begin{figure*}
\includegraphics[width=\linewidth]{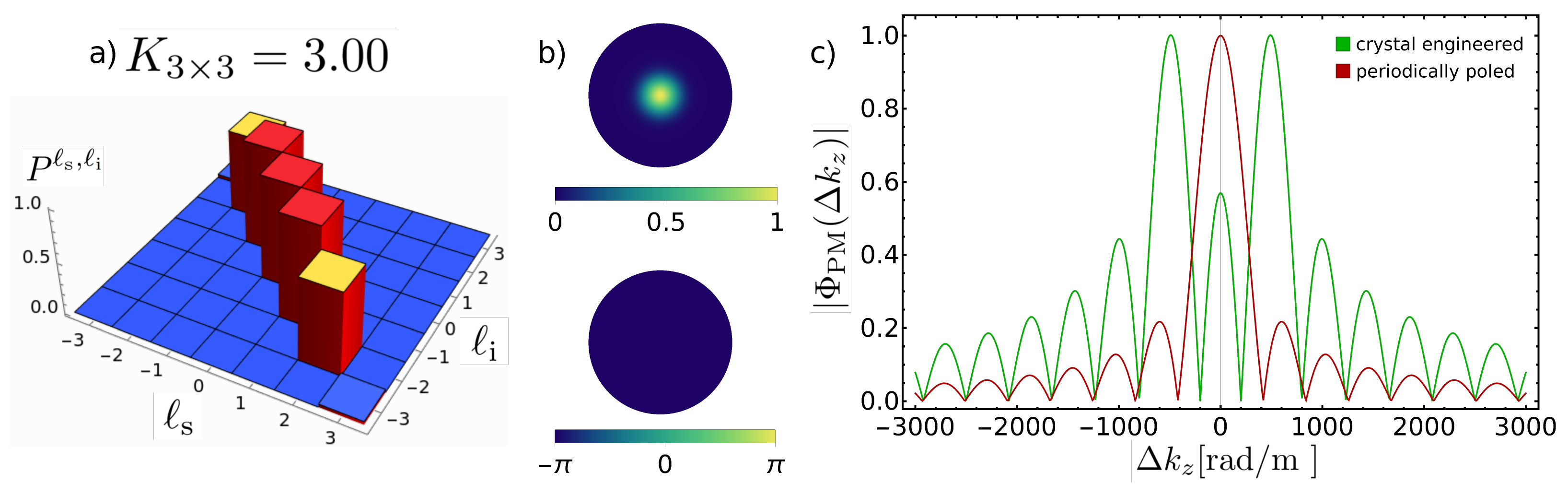}
\caption{\label{figure6}
a) The normalized OAM spectrum for the target state $\ket{\Psi_4}=\frac{1}{\sqrt{3}} \bigl( \ket{-1,1}+\ket{0,0}+\ket{1,-1} \bigl)$ within the range $\ell_{\signal}, \ell_{\idler}=-3, ..., 3$. The state is maximally entangled in the subspace $S_{3 \times 3}$. This example demonstrates that crystal engineering also enables to equalize OAM mode amplitudes. b) The pump is not a superposition of LG modes, but simply the fundamental Gaussian mode. c) The customized PMF $\Phi_{\text{PM}}(\Delta k)$ with $c_0=3.555$ and $c_1=-5.506$ (green) and for comparison, a typical sinc-like PMF of a periodically poled crystal (red).}
\end{figure*}

We extend our method to $d>3$ and investigate the generation of a two-ququint state ($d=5$) in the well-defined subspace $S_{5 \times 5}$. In this case, the subspace refers to $\mathcal{H}_{\signal} \otimes \mathcal{H}_{\idler}$ where $\mathcal{H}_{\signal} = \mathcal{H}_{\idler} = \bigl \{ \ket{-2},\ket{-1}, \ket{0}, \ket{1}, \ket{2}\bigl \}$. A neat example of combined pump and crystal engineering is shown in Fig. \ref{figure5} a) with the target state $\ket{\Psi_3}= \frac{1}{\sqrt{5}} \bigl( \ket{-2,-1} + \ket{-1,-2} + \ket{0,0} + \ket{1,2} + \ket{2,1} \bigl)$. Pump engineering enables to equalize the target mode amplitudes $C^{-1,-2}=C^{0,0}=C^{1,2}$. The conditions $C^{-1,-2}=C^{-2,-1}$ and $C^{1,2}=C^{2,1}$ are automatically fulfilled by degeneracy $k_{\signal}=k_{\idler}$. The modes generated by OAM conservation with non-zero projection probability of $\ell_{\pump}=-2$ and $\ell_{\pump}=2$ lie outside of the considered subspace. Thus, all target modes inside $S_{5 \times 5}$ have the RMN $N_R=0$ while the RMN of the unintended modes are $N_R=-2$ (modes $\ket{-1,1},\ket{1,-1}$) or $N_R=-4$ (modes $\ket{-2,2},\ket{2,-2}$). So we can apply crystal engineering to suppress the expansion amplitudes $C^{-2,2}=C^{-1,1}= C^{1,-1}=C^{2,-2}=0$ of the unintended modes. 

The intensity distribution of the pump beam features six peaks, see Fig. \ref{figure5} b). Given that $a_{-3}$ and $a_3$ are equally weighted, the calculated spatial profile is similar to Fig. \ref{figure4} b). The corresponding PMF in c) is determined by three crystal coefficients $c_0,c_1,c_2$. The number of crystal coefficients has increased because we now have more unintended modes to suppress than before. The calculated PMF shows two maxima enclosing three suppressed lobes around $\Delta k_z=0$.

We may note that sole pump engineering within the degenerate regime can achieve MESs within specific subspaces too. In $S_{3 \times 3}$, two examples could be $\frac{1}{\sqrt{3}} \bigl( \ket{-1, -1}+\ket{1,0}+\ket{0,1} \bigl)$ and $ \frac{1}{\sqrt{3}} \bigl(\ket{-1,0} + \ket{0,-1} +\ket{1,1} \bigl)$ for pump superpositions containing LG modes with $\ell_{\pump}=-2,1$ and $\ell_{\pump}=-1,2$, respectively. However, in other high-dimensional subspaces, we may equalize or suppress more than two modes along the same anti-diagonal. In this case, customized PMFs through crystal engineering prove to be highly advantageous.

\subsection{Crystal engineering with a Gaussian pump}

Finally, we examine a special case where only the crystal is engineered and the pump beam is left in the fundamental Gaussian mode. This means that the biphoton state is given by 
\begin{equation}
    ...+C^{-1,1}\ket{-1,1}+C^{0,0}\ket{0,0}+C^{1,-1}\ket{1,-1}+... \: .
\end{equation}
This state becomes maximally entangled if we can equalize the OAM amplitudes. We restrict the discussion to the subspace $S_{3 \times 3}$ and consider the target state $\ket{\Psi_4} = \frac{1}{\sqrt{3}} \bigl( \ket{-1,1} + \ket{0,0} + \ket{1,-1}  \bigl)$. Similar to Eq. \eqref{matrixforstate3.1}, but with $C^{0,0}=\frac{1}{\sqrt{3}}$ instead of $C^{1,-1}= 0$, we construct linear system of equations and find the solution $c_0=3.555$ and $c_1=-5.506$. 
The OAM spectrum is shown in Fig. \ref{figure6} a). Since all target modes lie along the anti-diagonal $\ell_{\pump}=0$, pump engineering is not required and the spatial pump profile corresponds to a Gaussian beam. The OAM spectrum without crystal engineering corresponds to Fig. \ref{figure1} a), whose Schmidt number is $K_{3 \times 3}=1.14$ indicating an almost separable state. The crystal-engineered state has $K_{3 \times 3}=3$ and is maximally entangled. This highlights that crystal engineering not only enables the suppression of OAM modes, but also allows to equalize the mode amplitudes. The customized PMF to achieve this OAM spectrum is displayed in green in Fig. \ref{figure6} c) and compared to the course to a periodically poled crystal (red). Similar to before, the calculated PMF has two symmetric maxima at $\Delta k_z \neq 0$ while the lobe at $\Delta k_z = 0$ is suppressed.

In principle, this MES can be generalized for odd $d>3$, like in $d=5$ to the MES $\frac{1}{\sqrt{5}} \bigl( \ket{-2,2} + \ket{-1,1} + \ket{0,0} + \ket{1,-1} + \ket{2,-2}  \bigl)$, where the calculated crystal coefficients lead to more sophisticated PMFs than in Fig. \ref{figure6} c).

\section{\label{sec:Conclusion}Conclusion}
In this paper, we have analyzed the entanglement of high-dimensional quantum states in OAM between photon pairs generated from SPDC.

We confirm that, although pump engineering methods can generate equally probable biphoton OAM modes, these do not represent high-dimensional maximally entangled states. OAM spectra generated solely by pump-engineered sources generally have unintended OAM modes present that obey the OAM conservation law for collinear SPDC. This results in a Schmidt numbers $K_{d \times d} < d$. We resolved this issue by demonstrating theoretically that suitable PMF modifications of the crystal can actually enhance the entanglement between signal and idler photons. We have presented PMF that enables the generation of MESs with $K_{d \times d} = d$. Our findings are relevant for implementing multidimensional quantum information processing applications, which aim to utilize genuine, well-defined two-qudit MES generated directly from SPDC without entanglement concentration. 

However, the use of a MES in a subspace requires the out-ruling of all modes that are not contained in the subspace. The extraction of target modes in the subspace could be achieved with quantum non-demolition (QND) measurements \cite{braginsky1980quantum, roch1992quantum, grangier1998quantum}, but their experimental realizations are very difficult. So far, the main interest was focused on the photon number \cite{brune1990quantum, chen2017quantum, wang2015quantum, yanagimoto2023quantum} or polarization \cite{sciarrino2006realization}, rather than OAM \cite{kaviani2020optomechanical, bierdz2011compact}. Alternatively, the amount of modes outside of the subspace may be reduced by a transition from the OAM basis to the basis of true Schmidt modes \cite{kovlakov2018quantum}. Otherwise, the modes belonging to the subspace still have to be post-selected. 

The future goal is to concentrate the entire flux of generated biphotons into the subspace, ensuring that higher-order modes have a negligible population. Fully controllable generation techniques should also consider spectral and radial DOFs of the LG modes to eliminate the need for post-selection.

\begin{acknowledgments}
The authors thank René Sondenheimer and Fabian Steinlechner for useful discussions regarding this work. 
\end{acknowledgments}

\appendix

\section{\label{appendixRMN} Details on the relative mode number}
\subsection{Advice to find tunable MES}
As discussed in the main text, the relative mode number (RMN) primarily determines the expansion amplitude in our discussion. If we want to suppress the probability of unintended modes, this only works if their RMN differs from the RMN of the target modes.

\begin{figure}[t]
\includegraphics[width=0.95\linewidth]{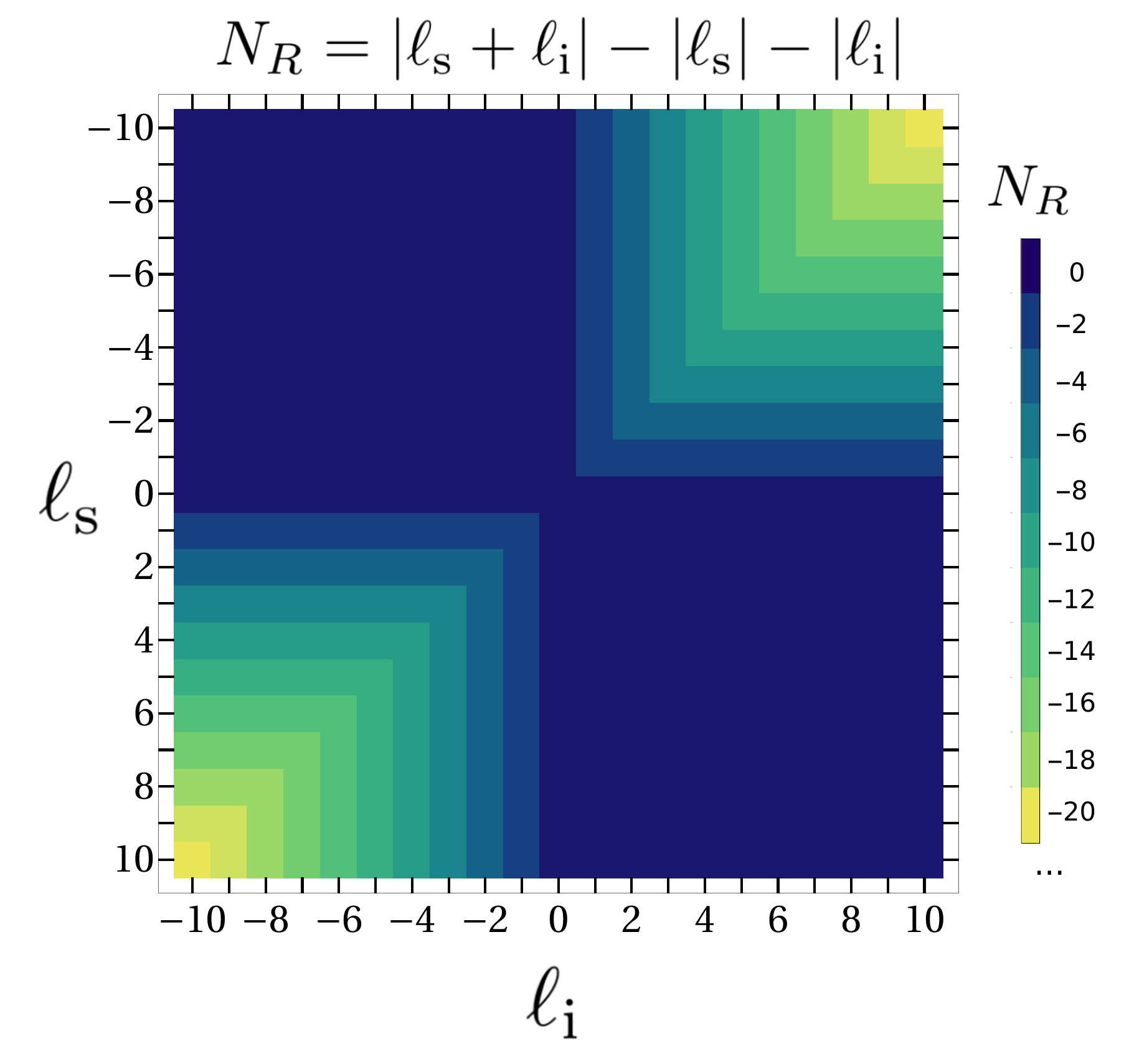}
\caption{\label{figure7} The relative mode number (RMN) $N_R = |\ell_{\signal}+\ell_{\idler}| - |\ell_{\signal}| -|\ell_{\idler}|$ of OAM modes $\ket{\ell_{\signal}, \ell_{\idler}}$ colorized within the range $\ell_{\signal}, \ell_{\idler}=-10, ..., 10$. Most modes have $N_R=0$ (dark blue). If either $\ell_{\signal}$ or $\ell_{\idler}$ is negative, the RMN can be a mutiple of $-2$. OAM modes of the same $N_R<0$ are grouped along vertical and horizontal strips. This pattern continues to infinity.}
\end{figure}

Fig. \ref{figure7} examines the RMN in more detail. By definition, most modes $\ket{\ell_{\signal}, \ell_{\idler}}$ have $N_R = 0$ (dark blue color). RMNs $N_R < 0$ in steps of negative even integers appear in a regular pattern only if either $\ell_{\signal}$ or $\ell_{\idler}$ is negative.
In detail, when fixing a signal OAM index $\ell_{\signal} \geq 0$, three horizontal areas can be distinguished when varying the idler OAM index: 
\begin{eqnarray*} 
\ell_{\idler} \geq 0: \: && N_R = 0 \\ 
-\ell_{\signal} < \ell_{\idler} < 0 : \: && N_R = -2, -4, \ldots, -2 |\ell_{\signal}| \\ 
\ell_{\idler} < -\ell_{\signal}: \: && N_R = -2 |\ell_{\signal}| \end{eqnarray*} 
For $\ell_{\signal} < 0$, the same applies, but with reversed inequality symbols. This behavior is also observed when swapping the roles of $\ell_{\signal}$ and $\ell_{\idler}$, except that vertical stripes of the same RMN can be seen.

We should construct MES in a subspace, in which different RMNs appear to effectively make use of crystal engineering. The most straightforward way is to generalize $S_{3 \times 3}$ to the subspace $ S_{d \times d}$, where $d$ is an odd number. Analogous, this denotes the two-photon Hilbert space $\mathcal{H}_{\signal} \otimes \mathcal{H}_{\idler}$ where $\mathcal{H}_{\signal} = \mathcal{H}_{\idler} = \bigl \{ \ket{-\frac{d-1}{2}},..., \ket{0}, ... , \ket{\frac{d-1}{2}} \bigl \}$. The subspace is symmetric around the mode $\ket{0,0}$. The dimension of the subspace $ S_{d \times d}$ is $d^2$. In this well-defined two-photon subspace MESs are characterized with an Schmidt number $K_{d \times d}=d$. The RMN in this subspace ranges from $N_R=0, -2, ..., -d+1$.

The RMN constraints compel us to choose MES with target modes, whose RMN are in the range $N_R=0,-2,...,m$ (with $m<0$), and the unintended modes created due to OAM conservation either
\begin{enumerate}[(i)]
    \item lie outside of the considered subspace $ S_{d \times d}$ if their RMN is $N_R=0,-2,...,m$ (i.e. we cut desirable OAM modes out) 
    or
    \item have a RMN  $N_R<m$.
\end{enumerate}
When these conditions are met, quantum state engineering for a MES with optimal crystal coefficients is possible.

\subsection{Determining suitable crystal coefficients}
In the subspace $S_{d \times d}$, we can tune only expansion amplitudes by crystal engineering that lie along a single anti-diagonal in the OAM spectrum. If we have several anti-diagonals, we would introduce separable modes or have several modes with $N_R=0$ (which all cannot be equalized by crystal engineering). We can equalize at most two modes with RMN $N_R=0$ along an anti-diagonal. 

Effectively, we are thus restricted to apply crystal engineering along $\ell_{\pump}=0$ (with expansion amplitude $C^{0,0}$), along $\ell_{\pump}=-1$ (with $C^{-1,0}=C^{0,-1}$) or along $\ell_{\pump}=1$ (with $C^{0,1}=C^{0,1}$). These conditions are automatically fulfilled due to degeneracy $k_{\signal}=k_{\idler}$.

Along one of these anti-diagonals, crystal engineering allows us to equalize (suppress) the amplitudes of the target (unintended) modes with different RMN. We find one condition for every RMN and their corresponding expansion amplitude. The expansion amplitude of all target modes have to be equal to $C^{\ell_{\signal},\ell_{\idler}} \propto 1/\sqrt{d}$ and expansion amplitude of unintended modes $C^{\ell_{\signal},\ell_{\idler}} \approx 0$. In the subspace $S_{d \times d}$, we find $\frac{d+1}{2}$ different RMNs. Due to the ansatz Eq. \eqref{CosineSus}, the conditions can directly be related to a linear system of $\frac{d+1}{2}$ equation to be solved for the crystal coefficients $c_n$, where $n$ is ranging from $0,...,\frac{d-1}{2}$:
\begin{widetext}
\begin{small}
\begin{equation*}
\frac{1}{\sqrt{d}} 
\begin{pmatrix}
1 \\[0.3em]
1 \\[0.3em]
\vdots \\[0.3em]
0 \\[0.3em]
\end{pmatrix} \overset{!}{=} 
\begin{pmatrix}
C^{\ell_s,\ell_i}({N_R=0}) \\[0.3em]
C^{\ell_s,\ell_i}({N_R=-2}) \\[0.3em]
\vdots \\[0.3em]
C^{\ell_s,\ell_i}({N_R=-d+1}) \\[0.3em]
\end{pmatrix} \overset{\eqref{RMNinExpansionAmplitude}}{\propto} 
\begin{pmatrix}
\int dz \: \xi(0,z) \cos{\Bigl( \frac{0 \cdot z}{\sigma} \Bigl)} & 
&
... & 
\int dz \:\xi(0,z) \cos{\Bigl( \frac{(d-1) z}{2 \sigma} \Bigl)}  \\[0.3em]
\int dz \:\xi(-2,z) \cos{ \Bigl( \frac{0  \cdot z}{\sigma} \Bigl)} & 
&
... & 
\int dz \: \xi(-2,z)\cos{\Bigl( \frac{(d-1) z}{2 \sigma} \Bigl)}  \\[0.3em]
 \vdots & 
&  \ddots
&
 \vdots  \\[0.3em]
\int dz \: \xi(-d+1,z) \cos{\Bigl( \frac{0  \cdot z}{\sigma} \Bigl)} & 
&
... &  
\int dz \: \xi(-d+1,z)  \cos{\Bigl( \frac{(d-1)z}{2 \sigma} \Bigl)}  \\[0.3em]
\end{pmatrix} 
\cdot
\begin{pmatrix}
c_0 \\
c_1 \\
\vdots \\
c_{\frac{d-1}{2}} \\
\end{pmatrix} 
\end{equation*}
\end{small}
\end{widetext}
The rows increase in steps of $-2$ from $N_R=0$ to $N_R=-d+1$, while the columns increase with the cosine argument in the range $n=0, 1, ...,\frac{d-1}{2}$. The number of $0$s and $1$s on the very left side depends on the explicit target state, i.e. how many modes have to be suppressed or equalized along the main anti-diagonal. By solving the system of equations for $\{c_0,...,c_{\frac{d-1}{2}} \}$, we have found the optimized crystal coefficients for the MES. 

\bibliography{apssamp}

\end{document}